\documentclass[preprint,prd,showpacs,showkeys,preprintnumbers,amsmath,amssymb]{revtex4}

\begin{document}

\preprint{APS/123-QED}

\title{Gravitational lensing in metric theories of gravity}

\author{Mauro Sereno}
\email{sereno@na.infn.it}
\affiliation{Dipartimento di Scienze Fisiche,
\\ Universit\`{a} degli Studi di Napoli ``Federico II"
\\ and
\\ Istituto Nazionale di Fisica Nucleare, Sez. di Napoli,
\\ via Cinthia, Compl. Univ. Monte S. Angelo,
\\ 80126, Napoli, Italia }

\date{December 6, 2002}

\begin{abstract}
\noindent Gravitational lensing in metric theories of gravity is discussed.
I introduce a generalized approximate metric element, inclusive of
both post-post-Newtonian (ppN) contributions and gravito-magnetic
field. Following Fermat's principle and standard hyphoteses, I derive
the time delay function and deflection angle caused by an isolated
mass distribution. Several astrophysical systems are considered. In
most of the cases, the gravito-magnetic correction offers the best
perspectives for an observational detection. Actual measurements
distinguish only marginally different metric theories one from
another.
\end{abstract}

\maketitle

\section{Introduction}
The principle of equivalence provides a firm foundation to any
conceivable theory of gravity. On the other hand, the derivation of
Einstein's field equations contains a strong element of guesswork. It
is, therefore, very interesting to test metric theories of gravity
defined as theories such that \cite{wil93,ci+wh95}: {\it i)}
space-time is a Lorentzian manifold; {\it ii)} the world lines of test
bodies are geodesics; {\it iii)} the equivalence principle in the
medium strong form is satisfied. General relativity, Brans-Dicke
theory and the Rosen bimetric theory satisfy these postulates. In
these theories, the usual rules for the motion of particles and
photons in a given metric still apply, but the metric may be different
from that derived from the Einstein's field equations. The basic
assumption of the existence of a dynamical space-time curvature, as
opposed to a flat space-time of special relativity, still holds.

Different metric theories can be compared with suitable tests. Bending
and time delay of electromagnetic waves are two important phenomena
predicted by theories of gravity. A comparison among general
relativity and other viable theories of gravity can be led on the
basis of higher-order effects. Intrinsic gravito-magnetism is such an
effect. Mass-energy currents relative to other masses generate
space-time curvature. In particular, the effect of the angular
momentum of the deflector has been studied by several authors with
very different approaches, see
\cite{ep+sh80,ib+ma82,iba83,dym86,gli99,as+ka00,io02pla,io02hom,se+ca02}.
The ppN corrections to the metric element have also to be considered.
These second order contributions, in the cases of either gravitational
deflection of light or relativistic time delay, have been discussed,
respectively, in \cite{ep+sh80,fi+fr80,ri+ma82a,ri+ma82b} and
\cite{ri+ma83}. In particular, using a parametrized expansion of the
metric element to second order in the Newtonian potential, Richter
\& Matzner \cite{ma+ri81,ri+ma82a,ri+ma82b,ri+ma83} have examined phenomena of
light deflection by Sun and Jupiter.

In this paper, I discuss deflection and time delay of light rays in
the usual framework of gravitational lensing \cite{pet+al01,sef}. As
shown in \cite{cap+al99,io02pla}, the standard assumptions of
gravitational lensing, i.e. the weak field and slow motion
approximation for the lens, allow us to consider higher-order
approximation terms in the calculation of lensing quantities, so that
a very general treatment of lensing phenomena can be performed.

On the observational side, gravitational lensing is one of the more
deeply investigated phenomena of gravitation and it is becoming a more
and more important tool for experimental astrophysics and cosmology.
The impressive development of technical capabilities makes it possible
to obtain observational evidences of peculiar metric theories of
gravity in a next future and to test the degree of accuracy of the
Einstein's field equations. Furthermore, observations of gravitational
lensing phenomena could demonstrate the inertia-influencing effect of
masses in motion. In fact, the gravito-magnetic field, predicted in
1896-1918, has not yet a firm experimental measurement.

The paper is as follows. In Section~\ref{gene}, I introduce the
generalized metric element in the weak field and slow motion
approximation. Section~\ref{ferm} considers the Fermat's principle in
stationary space-times. Sections~\ref{dela} and \ref{defl} contain a
derivation of, respectively, the time delay function and the
deflection angle. In Section~\ref{near}, I consider light rays passing
outside the lens; several astrophysical systems are discussed.
Section~\ref{summ} is devoted to a summary and to some final
considerations.

\section{A generalized metric element}
\label{gene}
The comparison of metric theories of gravity with each other and with
experiments can be performed in the slow-motion, weak field limit. Let
us consider the approximate ppN metric element generated by an
isolated mass distribution, with energy density $\rho$, negligible
pressure, and energy current density $j^i = \rho v^i$, $v \ll c$. As
shown in \cite{ma+ri81,ri+ma82a}, the knowledge of light propagation
to any given order requires knowledge of every component of the metric
to the same order. We first consider the generalized central mass
solution of general relativity (Schwarzchild metric) in isotropic
coordinates and expand it as power series in the small parameter
$\frac{G M}{r}$ to the ppN order. Then, we multiply the terms of this
expansion by dimensionless parameters. This expression can be
generalized to an arbitrary mass distribution by replacing $-\frac{G
M}{r}$ with the standard Newtonian potential $U$, solution of $\Delta
U = 4 \pi G \rho$, $U \ll c^2$. It is $U \sim {\cal O}(\varepsilon^2)$,
with $\varepsilon$ denoting the order of approximation. Finally, we
introduce the non-diagonal components of the metric tensor generated
by mass currents. We can write $g_{0i} \sim -4 V_i$, where $V_i$ is
the gravito-magnetic potential, solution of $\Delta V_i = 4 \pi G \rho
v_i$, $V_i \sim U v \sim {\cal O}( \varepsilon ^3)$. The final
expression for the approximate metric element is
\begin{equation}
\label{ppN1}
ds^2 \simeq \left[ 1+2\frac{U}{c^2}+2\beta\left( \frac{U}{c^2}
\right)^2 \right]c^2dt^2
- \left[ 1-2\gamma \frac{U}{c^2} +\frac{3}{2}\epsilon \left( \frac{U}{c^2} \right)^2\right]d{\bf x}^2
-8 \mu \frac{{\bf V {\cdot}}  d {\bf x}}{c^3} \  c dt.
\end{equation}
Asimptotically, the metric reduces to the Minkowski's one. $\beta$ and
$\gamma$ are two standard coefficients of the post-Newtonian
parametrized expansion of the metric tensor \cite{ci+wh95,wil93}.
$\beta$ is related to nonlinearity of mass contribution to the metric;
$\gamma$ measures space curvature produced by mass. In general
relativity, it is $\beta = \gamma =1$; in the Brans-Dicke theory,
$\beta=1$ and $\gamma =
\frac{1+\omega}{2+\omega}$. $\epsilon$ and $\mu$ are non standard
parameters. $\epsilon$ takes into account the ppN contribution to the
metric \cite{ep+sh80}; $\mu$ quantifies the contribution to the
space-time curvature of the mass-energy currents and measures the
strength of the intrinsic gravito-magnetic field \cite{ci+wh95}. In
general relativity, $\epsilon = \mu =1$. Additional terms in a
parametrized expansion of the metric element can also be considered
\cite{ma+ri81,ri+ma82a}.

The approximated metric element just introduced cannot describe every
conceivable metric theory of gravity. In particular, it does not
consider preferred frame effects, violations of conservation of four
momentum and preferred location effects\footnote{Even by including the
complete standard set of ten parameters, the post-Newtonian
parametrized expansion cannot include every conceivable metric theory
of gravity \cite{ci+wh95}.}. However, the metric in Eq.~(\ref{ppN1})
should be obeyed by most metric theories, with differences among them
occurring only in the numerical coefficients.

We will assume that during the time a light ray interacts with an
isolated distribution of matter, the configuration of that matter does
not change significantly. Then, the metric element can be considered
as stationary with
\begin{equation}
\label{wf2}
U ({\bf x}) \simeq -G \int_{\Re^3} \frac{\rho ({\bf x^{'}})}{ | {\bf
x} - {\bf x^{'}}|} d^3 x^{'},
\end{equation}
and
\begin{equation}
\label{wf3}
{\bf V} ({\bf x}) \simeq -G  \int_{\Re^3} \frac{(\rho {\bf v})({\bf
x^{'}})}{ | {\bf x} -{\bf x^{'}}|} d^3 x^{'}.
\end{equation}

Our approximation is accurate enough to encompass all astrophysical
tests that can be performed in the foreseeable future and makes it
possible to describe gravitational lensing in quite general metric
theories beyond the post-Newtonian order, without knowing any
particular field equation.

\section{The Fermat's principle in stationary space-times}
\label{ferm}
Within the assumptions of geometrical optics, Maxwell's equations have
approximate solutions in generic curved space-times \cite{sef}. From
the principle of equivalence, it turns out that light rays are null
geodesics. They can be characterized by the Fermat's principle which
states that a light ray from a source to an observer follows a
trajectory, from among all kinematically possible paths, whose arrival
time is stationary under first order variations of the path, $\delta
\tau =0 $. The Fermat's principle takes a version of particular
interest in stationary space-times \cite{sef,mar+al00,as+ka00}. We
consider a metric whose components $g_{\alpha \beta}$ are function of
the spatial coordinates $x^i$ only (roman indeces label spatial
coordinates). On a null curve, it is
\[
ds^2=g_{\alpha \beta}dx^{\alpha}dx^\beta=0;
\]
for the future-directed light ray,
\begin{equation}
\label{staz1}
c dt= -\frac{g_{i0}}{g_{00}}d x^i+\frac{d l_{\rm P}}{\sqrt{g_{00}}},
\end{equation}
where ${d l_{\rm P}}^2 \equiv \left( -g_{ij}
+\frac{g_{0i}g_{0j}}{g_{00}}\right)dx^i d x^j$ defines the spatial
metric \cite{la+li85}.

Let us consider an asymptotically flat space-time. The arrival time of
a light ray, whose spatial projection is $p$, to an asymptotic
observer is given by
\begin{equation}
\label{staz2}
t=\frac{1}{c}\int_p \frac{d l_{\rm P}}{\sqrt{g_{00}}}
-\frac{g_{i0}}{g_{00}}d x^i,
\end{equation}
and the Fermat's principle states
\begin{equation}
\label{staz3}
\delta \int_p n d l_{\rm P} =0,
\end{equation}
where the spatial paths $p$ are to be varied with fixed endpoints; $n$
is an effective index of refraction defined as
\begin{equation}
\label{staz4}
n \equiv -\frac{g_{i0}}{g_{00}} e^i+\frac{1}{\sqrt{g_{00}}},
\end{equation}
where $e^i \equiv \frac{d x^i}{d l_{\rm P}}$ is the unit tangent
vector of a ray. This version of the Fermat's principle is formally
identical with the classical one.

\section{The time delay function}
\label{dela}
Let us go, now, to apply the above results to our approximate metric
element. The proper arc length is
\begin{equation}
\label{delppN1}
d l_{\rm P} \simeq \left\{ 1-\gamma \frac{U}{c^2} +\left(
\frac{3}{4}\epsilon -\frac{\gamma^2}{2} \right)  \left( \frac{U}{c^2}
\right)^2  + {\cal O} (\varepsilon^6) \right\} d l_{\rm eucl},
\end{equation}
where $d l_{\rm eucl} \equiv \sqrt{ \delta_{ij}d x^i d x^j}$ is the
Euclidean arc length. Since
Eqs.~(\ref{staz2},~\ref{staz4},~\ref{delppN1}), the total travel time
reads
\begin{eqnarray}
\label{delppN2}
t & = & \int_p \ n \ d l_{\rm P} \\ & \simeq &\frac{1}{c}\int_p
\left\{ 1-(1+\gamma)\frac{U}{c^2}+
\left[
\frac{3}{2}-\beta+\gamma\left( 1-\frac{\gamma}{2} \right)+
\frac{3}{4}\epsilon \right]  \left( \frac{U}{c^2} \right)^2  +4\mu
\frac{V_i}{c^3}e^i \right\} d l_{\rm eucl}. \nonumber
\end{eqnarray}
The time delay of the path $p$ relative to the unlensed ray $p_0$ is
\begin{equation}
\label{wf6}
\Delta T \equiv \frac{1}{c}\left\{ \int_p n \ dl_{\rm P} - \int_{p_0} dl_{\rm P} \right\}.
\end{equation}
Equation~(\ref{wf6}) can be expressed as a sum of geometrical and
potential time delays
\[
\Delta T =\Delta T_{\rm geom}+\Delta T_{\rm pot}.
\]
The geometrical time delay,
\begin{equation}
\label{wf7}
c\Delta T_{\rm geom} \equiv \int_p \ d l_{\rm P} - \int_{p_0} \ d
l_{\rm P},
\end{equation}
is due to the extra path length relative to the unperturbed ray $p_0$.

The potential time delay $\Delta T_{\rm pot}$ is due to the
retardation of the deflected ray caused by the gravitational field of
the lens; it is defined as the difference between the total travel
time and the integral of the line element along the deflected path,
\begin{equation}
c\Delta T_{\rm pot} \equiv \int_p \ n\ d l_{\rm P} - \int_{p} \ d
l_{\rm P}.
\end{equation}

In usual lensing phenomena, the deflection angle of a light ray is
very small and we can treat the lens as thin \cite{sef,pet+al01}.
Since the deflection occurs essentially in a small region near the
deflector, the actual ray path can be approximated by combining its
incoming and outgoing asymptotes. This trajectory is a
piecewise-smooth null geodesics curve consisting of a null geodesic
from the source to the deflector and one from the deflector to the
observer. It is useful to employ the spatial orthogonal coordinates
$(\xi_1, \xi_2, l )$, centred on the lens and such that the $l$-axis
is along the incoming light ray direction ${\bf e}_{\rm in}$. The
three-dimensional position vector $\bf x$ can be expressed as ${\bf x}
= \mbox{\boldmath $\xi$} + l {\bf e}_{\rm in}$. The lens plane,
$(\xi_1,\xi_2)$, corresponds to $l=0$. With these assumptions, the
geometrical time delay is \cite{sef,pet+al01}
\begin{equation}
\label{wf9}
c\Delta T_{\rm geom} \simeq \frac{1}{2}\frac{D_{\rm d} D_{s}}{D_{\rm
ds}}\left|
\frac{\mbox{\boldmath $\xi$}}{D_{\rm d}}-\frac{\mbox{\boldmath $\eta$}}{D_{\rm s}} \right|^2,
\end{equation}
where $D_{\rm s}$ is the distance from the observer to the source,
$D_{\rm d}$ is the distance from the observer to the deflector and
$D_{\rm ds}$ is the distance from the deflector to the source;
$\mbox{\boldmath $\eta$}$ is the bidimensional vector position of the
source in the source plane.

The potential time delay can be written as the sum of three terms,
\begin{equation}
\label{potsum}
\Delta T_{\rm pot}=\Delta T_{\rm pot}^{\rm pN} +\Delta T_{\rm pot}^{\rm GRM} +\Delta T_{\rm pot}^{\rm ppN}.
\end{equation}
The first contribution contains the post-Newtonian correction to the
time delay. To calculate the contribution of order $G^{N}$ to the
lensing quantities, we need the path of the deflected light ray to the
order $G^{N-1}$ \cite{fi+fr80}. So, an integration along the line of
sight, ${\bf e}_{\rm in}$, is accurate enough to evaluate the pN
contribution. This corresponds to the Born approximation. It is
\begin{eqnarray}
c \Delta T_{\rm pot}^{\rm pN} & \equiv &
-\frac{(1+\gamma)}{c^2}\int_{\rm l.o.s.} U d l_{\rm eucl} \label{delaypN} \nonumber \\ &
\simeq & -2 (1+\gamma)
\frac{G}{c^2}\int_{\Re^2}d^2\xi^{'}\Sigma(\mbox{\boldmath $\xi$}^{'}){\rm ln}
\frac{|\mbox{\boldmath $\xi$} -\mbox{\boldmath $\xi$}^{'}|}{\xi_0}+\rm const \label{wf10a}
\end{eqnarray}
where $\Sigma$ is the projected surface mass density
\begin{equation}
\label{wf11}
\Sigma(\mbox{\boldmath $\xi$})\equiv \int_{\rm l.o.s.} \rho(\mbox{\boldmath $\xi$},l)\ dl,
\end{equation}
and $\xi_0$ is a scale-length in the lens plane.

The second contribution to the time delay derives from the
gravito-magnetic field,
\begin{equation}
\label{delayGRM}
c\Delta T_{\rm pot}^{\rm GRM} \equiv  \frac{4\mu}{c^3} \int_p  {\bf V}
{\cdot} {\bf e}\ d l_{\rm eucl}\ .
\end{equation}
Since the dragging of inertial frames induces a correction of order
$\varepsilon^3$, we can again assume the thin lens hypothesis
\cite{sef,pet+al01} and integrate Eq.~(\ref{delayGRM}) over the
unperturbed ray ${\bf e}_{\rm in}$. Is is easy to generalize the
results in \cite{io02pla}; we get
\begin{equation}
\label{wf10b}
c\Delta T_{\rm pot}^{\rm GRM} \simeq 8 \mu \frac{G}{c^3}\int_{\Re^2}
d^2\xi^{'}
\Sigma(\mbox{\boldmath $\xi$}^{'}) \langle {\bf v} {\cdot} {\bf e}_{\rm in} \rangle_{\rm l.o.s.} (\mbox{\boldmath $\xi$}^{'})
{\rm ln} \frac{|\mbox{\boldmath $\xi$} -\mbox{\boldmath
$\xi$}^{'}|}{\xi_0}+\rm const;
\end{equation}
$\langle {\bf v}{\cdot}{\bf e}_{\rm in}\rangle_{\rm l.o.s.}$ is the weighted
average, along the line of sight, of the component of the velocity
$\bf v$ orthogonal to the lens plane,
\begin{equation}
\label{wf12}
\langle {\bf v}{\cdot}{\bf e}_{\rm in}\rangle_{\rm l.o.s.} (\mbox{\boldmath $\xi$})\equiv
\frac{\int ({\bf v}(\mbox{\boldmath $\xi$},l){\cdot} {\bf e}_{\rm in})
\ \rho(\mbox{\boldmath $\xi$},l)\ dl}{\Sigma(\mbox{\boldmath $\xi$})}.
\end{equation}

The last term in Eq.~(\ref{potsum}) is the ppN one. The first
contribution to the ppN time delay derives from non-linear interaction
of matter with space-time, represented in the approximate metric
element, Eq.~(\ref{ppN1}), by terms which contain the square of the
Newtonian potential. We have
\begin{equation}
\label{delayppN}
c\Delta T_{\rm pot}^{\rm ppN (1)} \equiv \frac{1}{c^4} \left[
\frac{3}{2}-\beta+\gamma\left( 1-\frac{\gamma}{2} \right)+
\frac{3}{4}\epsilon \right] \int_p U^2 d l_{\rm eucl}
\end{equation}
Since the ppN order is ${\cal O}(G^2)$, the Born approximation breaks
down, so that we have to consider the integration of the Newtonian
potential on the deflected path, calculated at order ${\cal O}(G)$,
\begin{equation}
c\Delta T_{\rm pot}^{\rm ppN (2)} \equiv-\frac{(1+\gamma)}{c^2}
\left\{
\int_p U d l_{\rm eucl} - \int_{\rm l.o.s} U d l_{\rm eucl}
\right\}.
\end{equation}
Higher order corrections to the geometrical time delay also contribute
to the ppN time delay. The difference in the distances of closest
approach of the deflected and undeflected light rays represents a pN
quantity: first-order corrections in the calculation of this
difference are $\sim {\cal O}(G^2)$ and induce a ppN contribution to
the time delay. We will consider explicitly this third contribution in
the Section~\ref{near}. $\Delta T_{\rm pot}^{\rm ppN}$ cannot be
expressed in terms of an integral of elementary functions and
projected quantities, such as $\Sigma$.

We remind that the time delay function is not an observable, but the
time delay between two actual rays can be measured.

\section{The deflection angle}
\label{defl}
To derive the deflection angle, we apply the Fermat's principle,
Eq.~(\ref{staz2}). In the formalism of the previous sections, the
Fermat's principle can be restated as: actual light rays, given the
source position, are characterized by critical points of the total
time delay, i.e. $\Delta T (\mbox{\boldmath $\xi$})$ is stationary
with respect to variations of $\mbox{\boldmath $\xi$}$. The lens
equation is then obtained calculating
\begin{equation}
\label{wf16}
\nabla_{\mbox{\boldmath $\xi$}}\Delta T (\mbox{\boldmath $\xi$})=0;
\end{equation}
we get
\begin{equation}
\label{wf17}
\mbox{\boldmath $\eta$}=
\frac{D_{\rm s}}{D_{\rm d}}\mbox{\boldmath $\xi$}
-D_{\rm ds}\hat{\mbox{\boldmath $\alpha$}}(\mbox{\boldmath $\xi$});
\end{equation}
$\hat{\mbox{\boldmath $\alpha$}} \equiv - c \nabla_{\mbox{\boldmath
$\xi$}}
\Delta T_{\rm pot}$ is the deflection angle, i.e. the difference of the initial
and final ray direction.  Once again, the post-Newtonian and the
gravito-magnetic contribution to the deflection angle have a simple
expression. It is
\begin{equation}
\label{deflpN1}
\hat{\mbox{\boldmath $\alpha$}}^{\rm pN}(\mbox{\boldmath $\xi$} ) \simeq 2(1+\gamma)\frac{G}{c^2}\int_{\Re^2}d^2\xi^{'}\Sigma(\mbox{\boldmath $\xi$}^{'}) \frac{\mbox{\boldmath $\xi$} -\mbox{\boldmath $\xi$}^{'}}{|\mbox{\boldmath $\xi$} -\mbox{\boldmath $\xi$}^{'}|^2}.
\end{equation}
The equivalence principle, special relativity and Newtonian
gravitational theory imply that a photon must feel the gravity field
of  massive body. They yield only the ``1" part of the coefficient in
front of Eq.~(\ref{deflpN1}). This accounts for the deflection of
light relative to local straight lines. However, because of space
curvature, local straight lines are bent relative to asymptotic
straight lines. The contribution proportional to $\gamma$ in
Eq.~(\ref{deflpN1}) is just the bending due to the space metric,
described by the $g_{ii}$ components of Eq.~(\ref{ppN1}); $\gamma$
measures, at the post-Newtonian order, the curvature generated by an
isolated mass and varies from theory to theory.

The contribution of the gravito-magnetic field to the deflection angle
is
\begin{equation}
\label{deflpN2}
\hat{\mbox{\boldmath $\alpha$}}^{\rm GRM}(\mbox{\boldmath $\xi$} ) \simeq
-8 \mu\frac{G}{c^3}\int_{\Re^2}d^2\xi^{'}\Sigma(\mbox{\boldmath $\xi$}^{'})
\langle {\bf v}{\cdot}{\bf e}_{\rm in}\rangle_{\rm l.o.s.}(\mbox{\boldmath $\xi$}^{'})
\frac{\mbox{\boldmath $\xi$} -\mbox{\boldmath $\xi$}^{'}}{|\mbox{\boldmath $\xi$} -\mbox{\boldmath $\xi$}^{'}|^2}.
\end{equation}
The parameter $\mu$ tests intrinsic gravito-magnetism in conceivable
metric theories of gravity \cite{ci+wh95}. For an analysis of the
integral in Eq.~(\ref{deflpN2}), I refer to \cite{io02pla,se+ca02}.

The ppN contribution to the deflection angle has not a simple form in
terms of $\Sigma (\mbox{\boldmath $\xi$})$.

\section{The nearly point-like lens}
\label{near}

Let us consider light rays passing outside a matter distribution of
total mass $M$ and radius $R$. We will refer to this situation as to
the nearly point-like lens. If we place the origin of the coordinates
at the centre of mass of the system, the dipole term vanishes
identically \cite{bi+tr87}. The Newtonian potential, to order
$\varepsilon^4$, reads
\begin{equation}
U =-G \frac{M}{|{\bf x} |} + U_{J_2},
\end{equation}
where $U_{J_2}$ is the quadrupole term,
\begin{equation}
U_{J_2} \equiv J_2 G \frac{M}{|{\bf x}|^3} R^2\frac{3 \cos^2
\vartheta-1}{2};
\end{equation}
$J_2$ is the dimensionless coefficient of the second zonal harmonic
and $\vartheta$ is the angle between the symmetry axis and the field
point $\bf x$. Assuming the deflector to be symmetric about its
angular momentum vector ${\bf L}$, it is
\begin{equation}
\cos \vartheta = \frac{{\bf L} {\cdot} {\bf x}}{|{\bf L}||{\bf x}|}.
\end{equation}
The lensing quantities at the post-Newtonian order,
Eqs.~(\ref{delaypN},~\ref{deflpN1}), reduce to
\begin{equation}
c \Delta T_{\rm pot}^{\rm pN}=-2(1+\gamma )\frac{GM}{c^2}\ln \left(
\frac{
\xi}{\xi_0} \right),
\end{equation}
and
\begin{equation}
\hat{\mbox{\boldmath $\alpha$}}^{\rm pN}(\mbox{\boldmath $\xi$} )=
2(1+\gamma )\frac{GM}{c^2} \frac{ \mbox{\boldmath $\xi$}}{\xi^2}.
\end{equation}

The first higher order correction derives from the quadrupole moment.
Since $U_{J_2}$ is a higher-order term, we can integrate along the
unperturbed path. It is
\begin{eqnarray}
\int_p U_{J_2} dl & \simeq & \frac{1}{2} J_2 G M  R^2 \int_{-\infty}^{+\infty} \left[
\frac{3}{L^2} \frac{(L_{\rm l.o.s.} l+ L_1 \xi_1+ L_2 \xi_2)^2}{l^2+\xi^2} - \frac{1}{(l^2+\xi^2)^{3/2}}
\right] dl \nonumber \\
& = & - J_2 G M  R^2 \left[ 1-2(\hat{\mbox{\boldmath $\xi$}}{\cdot}\hat{\bf
L}_\perp)^2-\left(
\frac{L_{\rm l.o.s.}}{L} \right)^2 \right] \frac{1}{\xi^2};
\end{eqnarray}
$L$ is the modulus of the angular momentum; $L_{\rm l.o.s.}$ is the
projection along the line of sight; $L_1$ and $L_2$ are the components
on $\hat{\xi}_1$ and $\hat{\xi}_2$, respectively; $\hat{\bf L}_\perp
\equiv \left( \frac{L_1}{L}, \frac{L_2}{L},0 \right)$ is the
projection of the versor of the angular momentum on the lens plane.
The contribution to the time delay is, see also \cite{ri+ma83},
\begin{equation}
\label{delayquadru}
c \Delta T^{J_2}_{\rm pot} \simeq (1+\gamma) \frac{G M}{c^2}  J_2 R^2
\left[ 1-2(\hat{\mbox{\boldmath $\xi$}}{\cdot}\hat{\bf L}_\perp)^2-\left(
\frac{L_{\rm l.o.s.}}{L} \right)^2 \right] \frac{1}{\xi^2}.
\end{equation}
The predicted deflection due to the second zonal harmonic reads, see
also \cite{ep+sh80},
\begin{equation}
\hat{\mbox{\boldmath $\alpha$}}^{J_2}(\mbox{\boldmath $\xi$} ) \simeq  2(1+\gamma) \frac{G M}{c^2} J_2 R^2 \left\{
\left[ 1- 4(\hat{\mbox{\boldmath $\xi$}}{\cdot}\hat{\bf L}_\perp)^2- \left(
\frac{L_{\rm l.o.s.}}{L} \right)^2 \right] \hat{\mbox{\boldmath $\xi$}} + 2(\hat{\mbox{\boldmath $\xi$}}{\cdot}\hat{\bf L}_\perp)\hat{\bf L}_\perp \right\}
\frac{1}{\xi^3}.
\end{equation}
Circular symmetry is broken.

Let us now calculate the ppN correction. The first contribution comes
from those terms in the approximate metric element which are quadratic
in $U$. Since
\begin{equation}
\int_{\rm observer}^{\rm source} U^2 dl_{\rm eucl}
\simeq (G M)^2\int_{-\infty}^{+\infty}\frac{1}{\xi^2+l^2}dl= \pi \frac{(G M)^2}{\xi},
\end{equation}
it is
\begin{equation}
c \Delta T_{\rm ppN}^{(1)}=\pi \left[ \frac{3}{2}-\beta+\gamma\left(
1-\frac{\gamma}{2} \right)+ \frac{3}{4}\epsilon \right] \left(
\frac{GM}{c^2} \right)^2 \frac{ 1}{\xi}.
\end{equation}
The second contribution to the ppN time delay derives from the
integration of the Newtonian potential on the deflected path. Since
the linearized metric to the pN order is spherically symmetric, the
photon path lies in the plane through source, lens and observer. As
well known, the deflected trajectory at the pN order reads
\cite{fi+fr80,ri+ma83}
\begin{equation}
\Delta \xi \simeq (1+\gamma)\frac{G M}{c^2}\frac{1}{\xi}\left( l+(l^2+\xi^2)^{1/2} \right),
\end{equation}
where $\Delta \xi$ is the distance of the photon from an axis,
parallel to the optical axis, passing through $\mbox{\boldmath
$\xi$}$. On the deflected path, the modulus of the distance of the
field point reads
\begin{equation}
|{\bf x}| \simeq (l^2+\xi^2)^{1/2} \left[ 1 - \xi \frac{\Delta
\xi}{l^2+\xi^2}\right].
\end{equation}
We have
\begin{eqnarray}
\int_p U dl -\int_{\rm l.o.s} U dl
& \simeq & -(1+\gamma) \left( \frac{G M}{c} \right)^2
\int_{-\infty}^{+\infty} \left[ \frac{l}{(l^2+\xi^2)^{3/2}}+ \frac{1}{l^2+\xi^2} \right]dl \\
& = & -(1+\gamma) \left( \frac{G M}{c} \right)^2 \frac{\pi}{\xi};
\end{eqnarray}
the second contribution to the ppN time delay reads
\begin{equation}
c \Delta T_{\rm ppN}^{(2)}= (1+\gamma)^2 \left( \frac{G M}{c^2}
\right)^2
\frac{\pi}{\xi}.
\end{equation}
The last contribution to the ppN order comes from higher order
corrections in the estimate of the length of the deflected path. It is
\begin{eqnarray}
\int_p dl & \simeq & \int_{\Delta l} dl\left[ 1+\frac{1}{2}\left( \frac{d \Delta \xi}{d l} \right)^2 \right] \\
& = & \int_{\Delta l} dl +\frac{1}{2} (1+\gamma)^2 \left( \frac{G
M}{c^2}
\right)^2 \frac{1}{\xi^2} \int_{\Delta l} dl \left[ 1+2\frac{l}{(l^2+\xi^2)^{1/2}} + \frac{l^2}{l^2+\xi^2} \right];
\end{eqnarray}
integrating from $-D/2$ to $+D/2$, $D \gg \xi$, we get
\begin{equation}
\label{ppNgeom}
\int_p dl  \simeq D \left[ 1+ (1+\gamma)^2 \left( \frac{G M}{c^2} \right)^2 \frac{1}{\xi^2} \right]
- \frac{1}{2} (1+\gamma)^2 \left( \frac{G M}{c^2} \right)^2 \frac{\pi}{\xi}.
\end{equation}
The last term in the right hand side of Eq.~(\ref{ppNgeom}) does not
depend on $D$ and represents a local effect; it contributes to the ppN
time delay,
\begin{equation}
c \Delta T_{\rm ppN}^{(3)}= -\frac{1}2{}(1+\gamma)^2 \left( \frac{G
M}{c^2} \right)^2
\frac{\pi}{\xi}.
\end{equation}
Then, adding the three contributes, we get
\begin{equation}
\label{ppNdelay}
c \Delta T_{\rm pot}^{\rm ppN}= \pi \left[ 2 \left( 1+\gamma \right)
-\beta +
\frac{3}{4}\epsilon \right] \left(
\frac{GM}{c^2} \right)^2 \frac{ 1}{\xi} ,
\end{equation}
and
\begin{equation}
\label{ppNdefl}
\hat{\mbox{\boldmath $\alpha$}}^{\rm ppN}(\mbox{\boldmath $\xi$} )=
\pi \left[ 2 \left( 1+\gamma \right) -\beta + \frac{3}{4}\epsilon \right]
\left( \frac{GM}{c^2} \right)^2 \frac{ \hat{\mbox{\boldmath $\xi$} } }{\xi^2}.
\end{equation}
For $\beta$, $\gamma$ and $\epsilon$ of order of the unity, nearly
$80\%$ of the bending come from the standard post-Newtonian parameters
$\beta$ and $\gamma$, the remaining $20\%$ arises from the
non-standard $\epsilon$ coefficient. Our estimate agrees with the
earlier results in \cite{ep+sh80,fi+fr80,ri+ma82a}.

Eqs.~(\ref{ppNdelay},~\ref{ppNdefl}) can be easily generalized. For a
not specified mass distribution, we have
\begin{equation}
c \Delta T_{\rm pot}^{\rm ppN} \simeq  \left[ 2 \left( 1+\gamma
\right)-\beta + \frac{3}{4}\epsilon \right] \int_{\rm l.o.s.} \left( \frac{U}{c^2}\right)^2 d l_{\rm eucl} ,
\end{equation}
and
\begin{equation}
\hat{\mbox{\boldmath $\alpha$}}^{\rm ppN}(\mbox{\boldmath $\xi$} )\simeq -c \left[ 2 \left( 1+\gamma
\right)-\beta + \frac{3}{4}\epsilon \right]  \int_{\rm l.o.s.} \nabla_{\mbox{\boldmath $\xi$}} \left( \frac{U}{c^2}\right)^2 d l_{\rm eucl}.
\end{equation}

In order to compare the effect of dragging of inertial frames on the
deflection angle with the ppN contribution, we use the results for a
finite homogeneous sphere. For a deflector rotating about the
$\xi_2$-axis with angular momentum $L$, it is, outside the lens radius
\cite{io02hom,se+ca02},
\begin{equation}
\label{delaygrm}
\Delta T^{\rm GRM}(\xi, \theta)= -\mu \frac{4G}{c^3} L \frac{\cos \theta}{\xi} ,
\end{equation}
and
\begin{eqnarray}
\alpha_1^{\rm GRM}(\xi, \theta) & =& \mu \frac{4G}{c^3} \frac{ L}{\xi^2} \cos 2\theta, \label{hom1}   \\
\alpha_2^{\rm GRM}(\xi, \theta) & =& \mu \frac{4G}{c^3} \frac{ L}{\xi^2} \sin 2\theta, \label{hom2}
\end{eqnarray}
where $\xi$ and $\theta$ are the polar coordinates in the lens plane.
The gravito-magnetic field breaks the circular symmetry. Both the ppN
and the gravito-magnetic contributions to the deflection angle
decrease as $\xi^{-2}$. The sign of the gravito-magnetic correction
varies on opposite sides of the lens, so that it can be separated
experimentally from other terms.

For an angular momentum directed along the $\xi_2$-axis, the
quadrupole contribution to the time delay, Eq.~(\ref{delayquadru}),
reduces to
\begin{equation}
\label{delayquadrubis}
c \Delta T^{J_2}_{\rm pot} \simeq (1+\gamma) \frac{G M}{c^2}  J_2 R^2
\frac{\cos 2 \theta}{\xi^2}.
\end{equation}
As can be seen comparing Eqs.~(\ref{delaygrm},~\ref{delayquadrubis}),
the angular dependence of the gravito-magnetic correction differs from
that in Eq.~(\ref{delayquadrubis}). The sign of the quadrupole
correction does not change in the equatorial plane. Furthermore, for a
deflector with angular momentum with generic direction in the space,
the gravito-magnetic effect depends only on the component of $\bf L$
in the lens plane \cite{io02pla}, whereas $L_{\rm l.o.s.}$ enters the
quadupole correction.

The magnitudes of the different contributions to the deflection angle
are considered by investigating real astrophysical systems acting as
lenses. It is enough to use the values of the coefficient in general
relativity, $\beta = \gamma = \epsilon = \mu =1$. We will consider
light rays in the equatorial plane ($\theta =0$).

The post-Newtonian deflection angle for rays grazing the solar limb is
1.75 arcsec; $\alpha^{\rm ppN}$ is about 11 $\mu$arcsec, where the
contribution of the non-standard $\epsilon$ coefficient is $\sim 2$
$\mu$arcsec. Given the angular momentum of the Sun, $L_{\odot}\simeq
1.6{\times} 10^{48}$g~cm$^2$s$^{-1}$  \cite{all83}, the gravito-magnetic
correction is $\sim 0.7$ $\mu$arcsec. Very Long Baseline
Interferometry (VLBI) has improved the accuracy of the measurements of
the deflection of radio waves by the Sun to the milli-arcsec level.
This is not enough to measure the higher order ppN and
gravito-magnetic contributions, so that the parameters $\beta$,
$\epsilon$ and $\mu$ cannot be determined. However, strong constraints
on $\gamma$ can be put. It is $\gamma = 1.000 \pm 0.002$
\cite{rob+al91}, an impressive confirmation of the prediction by
general relativity. In Brans-Dicke theory, this measurement constrains
the $\omega$ parameter, $\omega \stackrel{>}{\sim} 500$.

For an early type star,
$L=10^2L_{\odot}\left(\frac{M}{M_{\odot}}\right)^{5/3}$ \cite{kra67}.
For $M=1.4 M_{\odot}$, $R=1.1R_{\odot}$ and for a light ray grazing
the star's limb, $\alpha^{\rm pN} \simeq 2.23$ arcsec, $\alpha^{\rm
ppN}
\simeq 18$ $\mu$arcsec, $\alpha^{\rm GRM} \simeq 0.10$ milli-arcsec. The
gravito-magnetic correction overwhelms the ppN one by an order of
magnitude.

The gravito-magnetic field becomes even more significant for a fast
rotating white dwarf, where $L \sim \sqrt{0.2 G M^3 R}$ \cite{pad99}.
For $M \sim M_\odot$, $R \sim 10^{-2}R_\odot$, $\xi
\sim 6 R$, it is $\alpha^{\rm pN} \simeq 29.2$ arcsec, $\alpha^{\rm ppN} \simeq 0.1$
milli-arcsec, and $\alpha^{\rm GRM} \simeq 0.032$ arcsec. In this
case, the gravito-magnetic correction is quite important.

Now, we want to apply our approximation to a galaxy acting as a lens.
We take $M=10^{12}M_{\odot}$, $R \simeq 50$ kpc and $L \sim 0.1
M_{\odot}$ kpc$^2$s$^{-1}$, as derived from numerical simulations
\cite{vit+al01}. It is $\alpha^{\rm pN} \simeq 0.80$ arcsec,
$\alpha^{\rm ppN} \simeq 2.2$ $\mu$arcsec, $\alpha^{\rm GRM}
\simeq 0.16$ milli-arcsec. The gravito-magnetic correction overwhelms the
ppN one by two orders of magnitude.

\section{Summary}
\label{summ}
The time delay function and the deflection angle for a single lens
plane have been considered in the framework of metric theories of
gravity. I used an approximate metric element generated by an isolated
mass distribution in the weak field regime and slow motion
approximation, expanded to the ppN order, and with non-diagonal
components which include the effects of gravity by currents of mass.

Simple expressions for the ppN corrections to the lensing quantities
have been derived. The nearly point-like lens has been used to
consider several astrophysical systems. This approximation for the
deflector is quite rough, but, some of the times, astrophysics can be
tough. The gravito-magnetic correction and the ppN contribution to the
deflection angle are of the same order for intermediate main sequence
stars, like the Sun. For early type stars, white dwarfs and galaxies
acting as lenses, the gravito-magnetic term overwhelms the ppN one.
Differently from both pN and ppN contributions, which depend only on
the impact parameter of the incident ray, the gravito-magnetic field
induces a deflection which breaks the spherically symmetry, so that it
offers the best perspectives for an observational detection. However,
a mass quadrupole moment can make the situation worse on the
observational side.

Ground based instrumentations, such as VLBI, or satellites, such as
Hipparcos, can measure deflection angles, respectively in the
radio-wave regime and optical band, with accuracy of nearly
milli-arcsec. Since the $\gamma$ parameter appears in the
post-Newtonian expression of the lensing quantities, this accuracy put
strong constraints on it. However, the other parameters which enter
the approximate metric element, that is the standard $\beta$ term and
the non-standard $\epsilon$ and $\mu$ coefficients, need more accurate
measurements. Lensing by fast rotating stars, such as white dwarfs,
could give some hints on the dragging of inertial frames, whose
strength is determined by the $\mu$ parameter.

New generation space interferometric mission, such as SIM by NASA
(scheduled for launch in 2009), should greatly improve the
experimental accuracy, so that gravitational lensing could address, in
the near future, two very interesting topics in gravitation: the
detection of gravito-magnetism and possible discrepancies of gravity
from predictions of general relativity.

\begin{acknowledgments}
I thank the referee for the illuminating report.

\end{acknowledgments}

\end{document}